\documentclass[aps,prc,twocolumn,superscriptaddress]{revtex4-2}
\usepackage{xcolor}
\usepackage{graphicx}
\usepackage{subfig}    % 引入子图表包
\usepackage{overpic}
\usepackage{caption}
\usepackage{booktabs} 
\usepackage{enumitem}
\usepackage{amsmath}
\usepackage{multirow}
\usepackage{makecell}
\begin{document}

%\preprint{}

\title{Revisiting identified-particle $p_{\mathrm{T}}$ spectra using the Boltzmann–Gibbs blast-wave model in a Bayesian inference framework}


\author{Z. Xie}
\author{W.Z. Li}
\affiliation{School of Physics and Information Technology, Shaanxi Normal University, Xi'an 710119, China}

%\thanks{}
\author{J.Q. Tao}
\affiliation{Key Laboratory of Quark and Lepton Physics (MOE) and Institute of Particle Physics, Central China Normal University, Wuhan 430079, China}
\affiliation{School of Science and Engineering, The Chinese University of Hong Kong, Shenzhen (CUHK-Shenzhen), Guangdong, 518172, China}

\author{H. Zheng}
\email[]{zhengh@snnu.edu.cn}

\author{W.C. Zhang}
\affiliation{School of Physics and Information Technology, Shaanxi Normal University, Xi'an 710119, China}

\author{W. Dai}
\affiliation{School of Mathematics and Physics, China University of Geosciences, Wuhan 430074, China}

\author{L.L. Zhu}
\affiliation{College of Physics, Sichuan University, Chengdu 610064, China}

\author{X.Q. Liu}
\affiliation{Institute of Nuclear Science and Technology, Sichuan University, Chengdu 610064, China}

\author{D.M. Zhou}
% \email[]{zhoudm@mail.ccnu.edu.cn}
\affiliation{Key Laboratory of Quark and Lepton Physics (MOE) and Institute of Particle Physics, Central China Normal University, Wuhan 430079, China}

\author{B.H. Sa}
% \email[]{sabhliuym35@qq.com}
\affiliation{Key Laboratory of Quark and Lepton Physics (MOE) and Institute of Particle Physics, Central China Normal University, Wuhan 430079, China}
\affiliation{China Institute of Atomic Energy, P. O. Box 275 (10), Beijing 102413, China}


\date{\today}

\begin{abstract}
We perform a Bayesian analysis of transverse momentum ($p_{\mathrm{T}}$) spectra of identified particles, i.e., pions, kaons, and protons, at midrapidity in Au+Au collisions and Pb+Pb collisions using the Boltzmann–Gibbs blast-wave (BGBW) model. We investigate whether it is possible to simultaneously describe the $p_{\mathrm{T}}$ spectra of identified particles without imposing the particle species-dependent $p_{\mathrm{T}}$ fit ranges -- a practice that was followed in conventional blast-wave model studies to achieve reasonable simultaneous fits. Using Bayesian analysis, our results indicate that a simultaneous description of the $p_{\mathrm{T}}$ spectra of pions, kaons, and protons is feasible without imposing the particle species-dependent $p_{\mathrm{T}}$ fit ranges, for Au+Au collisions up to the available data ($\sim$2 GeV/c) and for Pb+Pb collisions up to 3 GeV/c. The extracted parameters remain broadly consistent with those obtained from conventional BGBW simultaneous fits, while the extension of the fit range leads to moderate changes in some parameters. Furthermore, Bayesian analysis yields well-constrained posterior distributions for the kinetic freeze-out temperature $T_{kin}$, the average transverse flow velocity $\langle \beta_{\mathrm{T}}\rangle$, and the exponent of the velocity profile $n$ and shows their correlations transparently. We suggest that the BGBW model in a Bayesian inference framework proposed can be applied in future data analyses to simultaneously describe the $p_{\mathrm{T}}$ spectra of identified particles and extract the relevant information about the collision system.
\end{abstract}

%\keywords{}

\maketitle

\section{INTRODUCTION}{\label{intro}}
Quark-Gluon Plasma (QGP), a state of deconfined quarks and gluons, is formed in high-energy heavy-ion collisions at the Relativistic Heavy-Ion Collider (RHIC) and the CERN Large Hadron Collider (LHC). The created matter exhibits strong collective flow, behaving as a nearly perfect liquid with an extremely low shear viscosity-to-entropy-density ratio $\eta/s$ \cite{Huovinen:2006jp,Muller:2006ee}. A primary goal of relativistic heavy-ion collisions is to study the properties of the QGP as well as the dynamics of the ``little bang." The transverse momentum ($p_{\mathrm{T}}$) distribution of identified hadrons carries information about both the thermal state at kinetic freeze-out and the collective radial flow developed during the expansion, serving as a probe to investigate the evolution of the system \cite{STAR:2003jwm,STAR:2008med,STAR:2017sal,STAR:2019vcp,STAR:2019bjj,STAR:2006xud,STAR:2006nmo,ALICE:2014juv,ALICE:2015ial,ALICE:2015qqj,ALICE:2011gmo,ALICE:2022kol,ALICE:2013mez,ALICE:2013cdo,ALICE:2019hno}. 
 
Many models based on hydrodynamics have been developed to describe the transverse momentum distributions of particles, including relativistic hydrodynamic models \cite{Osada:2008sw,Kyan:2022eqp,Shi:2024pyz}, the Boltzmann–Gibbs blast-wave (BGBW) model \cite{Schnedermann:1993ws,Schnedermann:1994gc,Chen:2020zuw}, and the Tsallis blast-wave model (TBW) \cite{Tang:2008ud,Shao:2009mu,Tang:2011xq,Ristea:2013ara,Zheng:2015gaa,Che:2020fbz,Gu:2022xjn}. The blast-wave model is a simplified hydrodynamic model. It assumes that a locally thermalized medium expands with a radial flow velocity $\beta$ and then undergoes an instantaneous kinetic freeze-out at the temperature $T_{kin}$. By fitting the $p_{\mathrm{T}}$ spectra of identified particles across various collision centralities and energies, one can extract key parameters such as the kinetic freeze-out temperature and the average radial flow velocity, which characterize the system at kinetic freeze-out, and study their evolution as a function of centrality and/or collision energy  \cite{Werner:2024ntd,Waqas:2022omn,Tao:2022tcw,Zhu:2022dlc,Zhu:2022bpe,Zhu:2021fbs,Zhu:2018nev,Zheng:2015gaa,Zheng:2015tua,Wong:2015mba,Rath:2019cpe,Xu:2017akx,Shi:2024pyz}. Among them, the BGBW model \cite{Schnedermann:1993ws,Schnedermann:1994gc,Chen:2020zuw} is widely adopted to describe the $p_{\mathrm{T}}$ spectra of identified particles at kinetic freeze-out by both experimentalists and theorists because of its successful applications in a wide range of collision systems and energies. These include Au+Au collisions at $\sqrt{s_{\mathrm{NN}}}$ = 7.7–200 GeV \cite{STAR:2003jwm,STAR:2008med,STAR:2017sal,STAR:2019vcp,STAR:2019bjj,PHENIX:2013kod,Tariq:2024hfc,STAR:2007zea}, $d$+Au collisions at $\sqrt{s_{\mathrm{NN}}}$ = 200 GeV \cite{STAR:2008med}, Pb+Pb collisions at $\sqrt{s_{\mathrm{NN}}}$ = 2.76 and 5.02 TeV \cite{ALICE:2013mez,ALICE:2013cdo,ALICE:2019hno,Tariq:2024hfc}, and $p$+Pb collisions at $\sqrt{s_{\mathrm{NN}}}$ = 5.02 TeV \cite{ALICE:2013wgn}. Furthermore, the BGBW model has been extended to describe high-multiplicity $p+p$ collisions \cite{Ghosh:2014eqa}. 

However, the Boltzmann–Gibbs blast-wave model was adopted to fit $p_{\mathrm{T}}$ spectra of identified particles with species-dependent fit ranges to achieve reasonably good simultaneous fits in practice, such as 0.5–1 GeV/c for pions, 0.2–1.5 GeV/c for kaons, and 0.3–3 GeV/c for protons in Pb+Pb collisions respectively, and then to extrapolate to the fit $p_{\mathrm{T}}$ region beyond in literature \cite{ALICE:2013mez,ALICE:2019hno}. Similar selected fit ranges have also been applied in other collision systems, including Au+Au collisions at RHIC \cite{STAR:2003jwm,STAR:2008med,STAR:2017sal,STAR:2019vcp}, $p$+Pb collisions \cite{ALICE:2013wgn}, and high-multiplicity $p+p$ collisions \cite{Ghosh:2014eqa}. These selected $p_{\rm T}$ fit ranges were empirically chosen with physical considerations to minimize contributions from resonance decays at low $p_{\mathrm{T}}$ and hard processes at high $p_{\mathrm{T}}$, in particular for pions \cite{Schnedermann:1993ws}. A natural question arises whether it is necessary to enforce the particle species-dependent $p_{\mathrm{T}}$ fit ranges when the BGBW model is applied to describe the $p_{\mathrm{T}}$ spectra of identified particles simultaneously and still achieve reasonably good fits. It is also interesting to see how the extracted parameters vary compared to the conventional fits in literature.

In this work, we revisit the $p_{\mathrm{T}}$ spectra of identified particles in Au+Au collisions at $\sqrt{s_{\mathrm{NN}}} = 7.7-200$ GeV and in Pb+Pb collisions at $\sqrt{s_{\mathrm{NN}}} = 2.76$ and $5.02$ TeV using the Boltzmann–Gibbs blast-wave model in a Bayesian inference framework. We address the question of whether it is possible to simultaneously describe the $p_{\mathrm{T}}$ spectra of identified particles without imposing the particle species-dependent $p_{\mathrm{T}}$ fit ranges -- a practice that was followed in conventional blast-wave model studies. We also investigate how the extracted parameters vary when changing the selected $p_{\mathrm{T}}$ fit ranges for different particle species. We note that there are two recent Bayesian analyses using blast-wave model \cite{Saha:2025nyu, Vitiuk:2024cor}, but with different physics goals. Ref.~\cite{Saha:2025nyu} focused on the transverse momentum differential radial flow observable $v_{0}(p_{\mathrm{T}})$, while Ref.~\cite{Vitiuk:2024cor} studied the low-$p_{\mathrm{T}}$ enhancement of the particle spectra for pions and kaons by introducing the nonequilibrium chemical potentials as an alternative to the conventional explanation, i.e., attributed to resonance decays with subsequent thermalization. Bayesian analysis can yield full posterior distributions of the model parameters, offering a more complete characterization of parameter uncertainties and correlations. This enables us to examine the validity of the BGBW model description over a broader kinematic region and to assess the sensitivities and correlations of the extracted model parameters.

The paper is organized as follows. Section~\ref{mod} provides a brief introduction to the Boltzmann–Gibbs blast-wave model and the Bayesian inference approach. In Sec.~\ref{rad}, we present the main results, including the transverse momentum spectra of $\pi$, $K$ and $p$ in Au+Au collisions and Pb+Pb collisions from the BGBW model in the Bayesian inference framework, their comparison with experimental data, and posterior distributions of the model parameters. A summary is given in Sec. \ref{con}.

\section{The BGBW model and Bayesian inference approach}\label{mod}
\subsection{The BGBW model}
The BGBW model is based on the assumption of local thermal equilibrium using the Boltzmann-Gibbs distribution. In the BGBW model, the invariant particle distribution \cite{Schnedermann:1993ws,Schnedermann:1994gc} is given by
\begin{equation}
    E\frac{d^3N}{dp^3} \propto \int_{0}^{R} m_{\mathrm{T}}I_0\left(\frac{p_{\mathrm{T}}\sinh\rho}{T_{kin}}\right)K_1\left(\frac{m_{\mathrm{T}}\cosh\rho}{T_{kin}}\right)rdr,
\end{equation}
where the velocity profile $\rho$ is parametrized as
\begin{equation}
    \rho=\mathrm{tanh^{-1}}\beta_{\mathrm{T}}(r)=\mathrm{tanh^{-1}}[(\frac{r}{R})^n\beta_s].
\end{equation}
\( m_{\mathrm{T}} = \sqrt{p_{\mathrm{T}}^2 + m^2} \) is the transverse mass, \(p_{\mathrm{T}}\) is the transverse momentum, \(m\) is the rest mass of the particle, and \( I_0 \) and \( K_1 \) are the modified Bessel functions. \(T_{kin}\) is the temperature at kinetic freeze-out. \( r \) and \( R \) represent the radial distance from the origin in the transverse plane and the fireball radius, respectively. \( \beta_{\mathrm{T}} \) is the transverse expansion velocity and \( \beta_s \) is the transverse expansion velocity at the surface, and \(n\) is the exponent of the velocity profile.  The average radial expansion velocity \( \langle \beta_{\mathrm{T}} \rangle \) is related to \( \beta_s \) \cite{Ristea:2013ara} by 
\[\langle \beta_{\mathrm{T}} \rangle = \frac{\int_0^R \beta_{\mathrm{T}}(r) r \, dr}{\int_0^R r \, dr},\ \beta_s = \langle \beta_{\mathrm{T}} \rangle \frac{n+2}{2}.\]  
\begin{figure*}[!ht]
    \centering
    \subfloat{
	\begin{overpic}[width=0.48\linewidth]{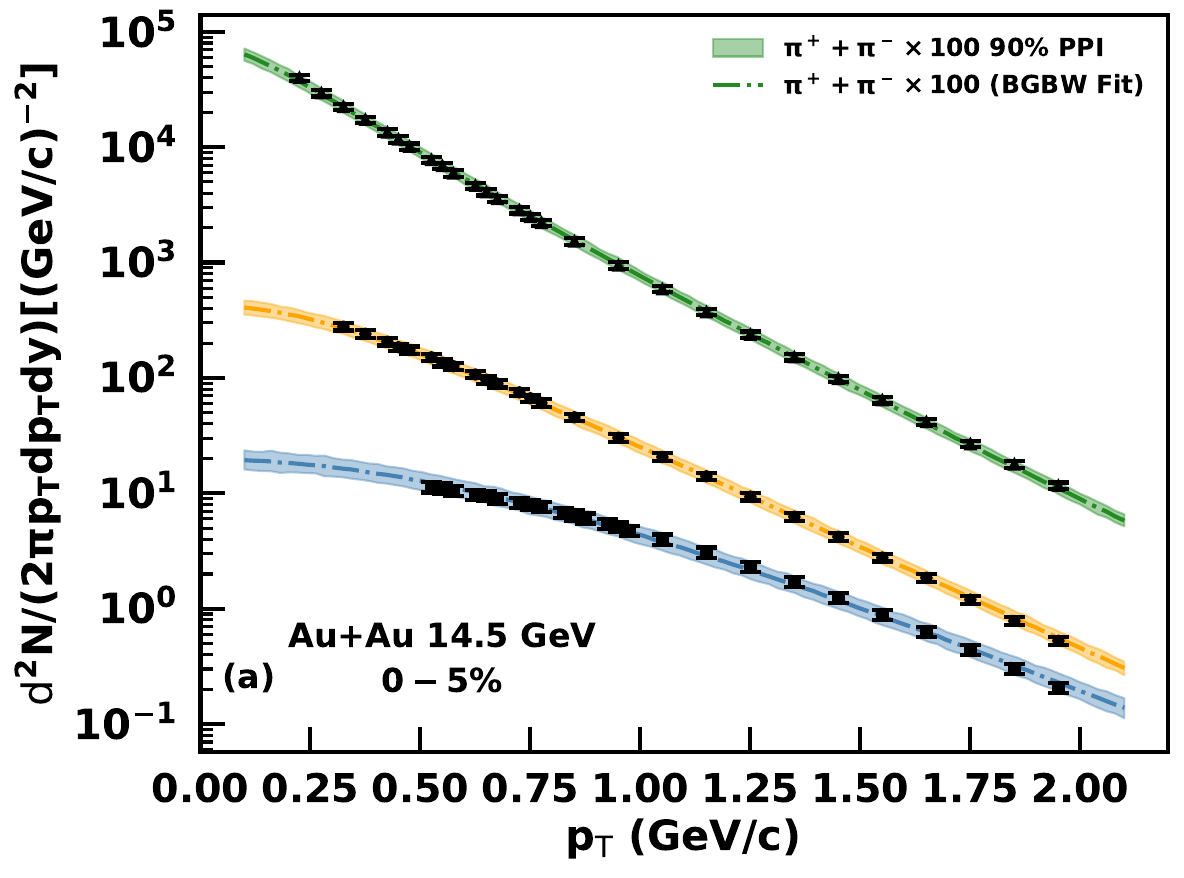}
    \end{overpic}
	}
    \hfill
     \subfloat{
	\begin{overpic}[width=0.48\linewidth]{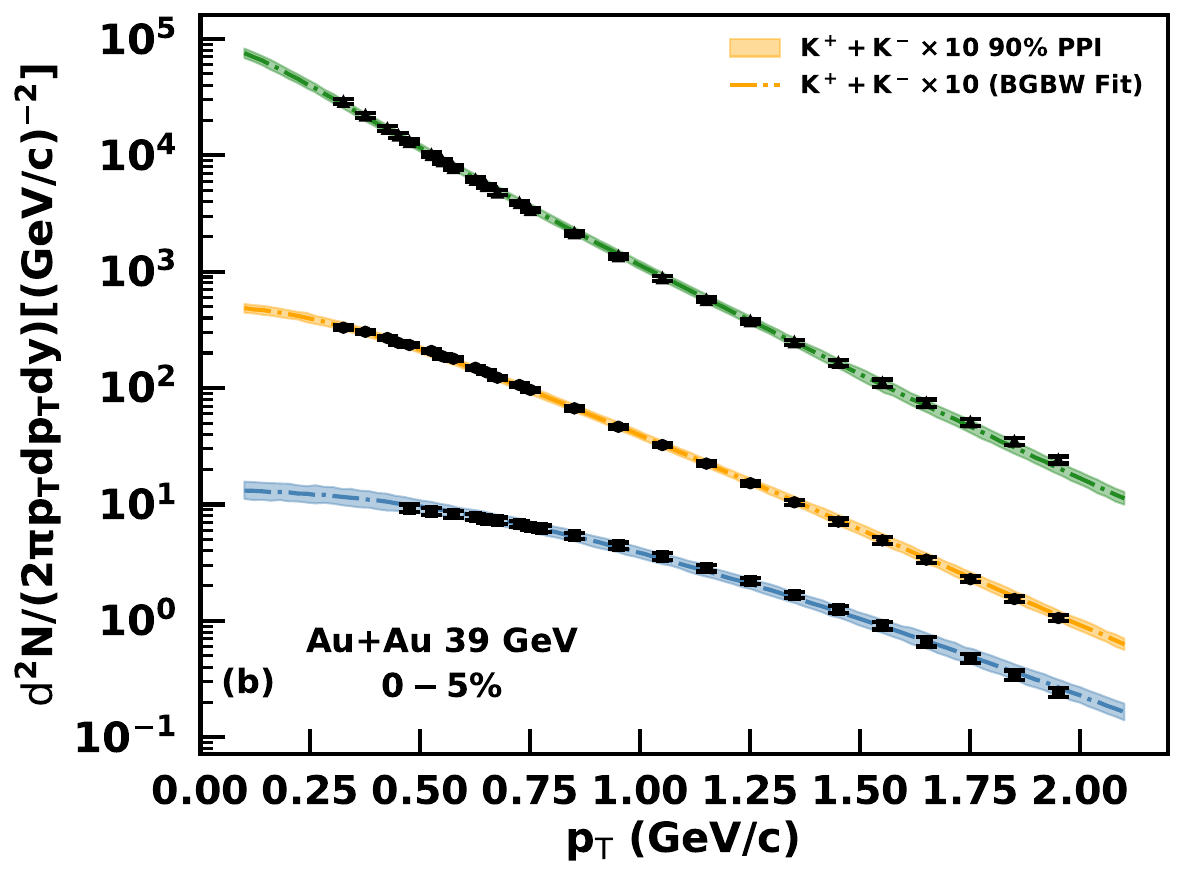}
    \end{overpic}
	}
    \hfill
    \subfloat{
	\begin{overpic}[width=0.48\linewidth]{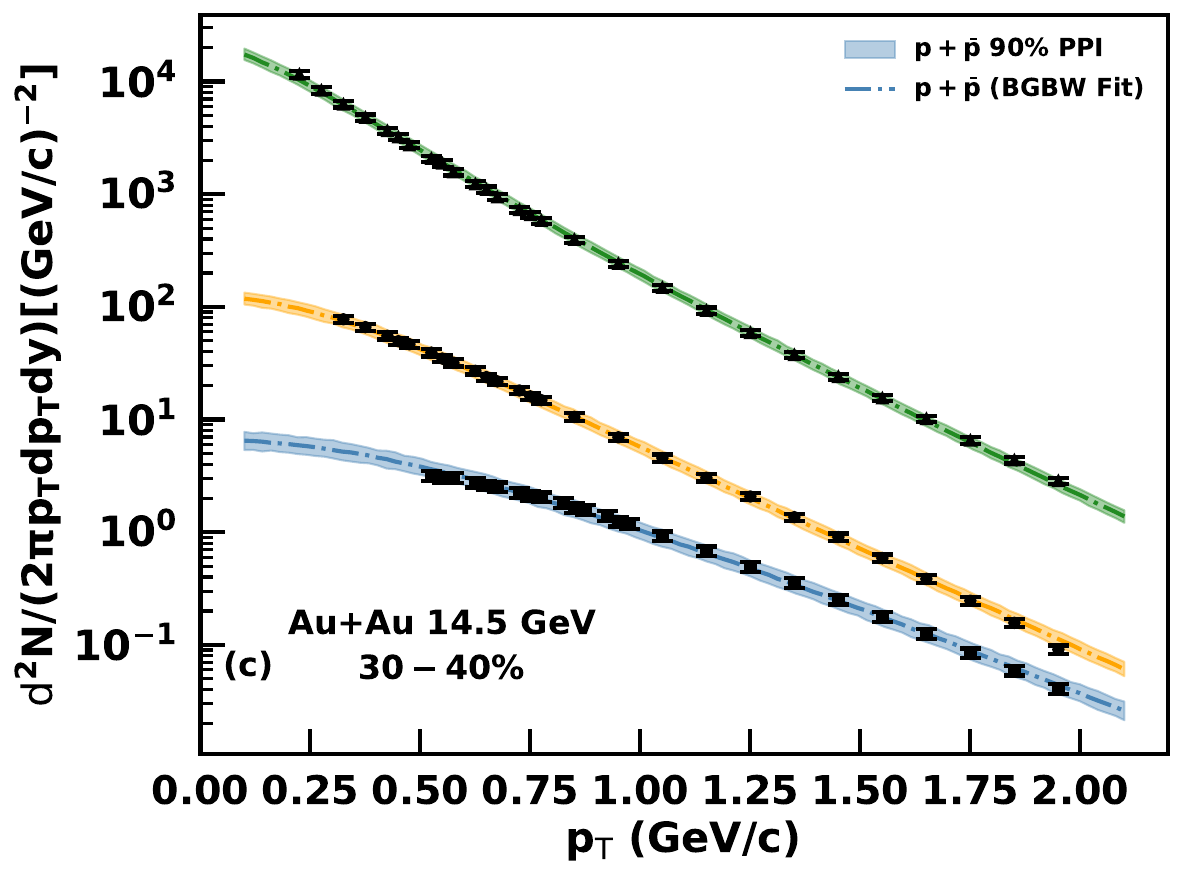}
    \end{overpic}
	}
    \hfill
    \subfloat{
	\begin{overpic}[width=0.48\linewidth]{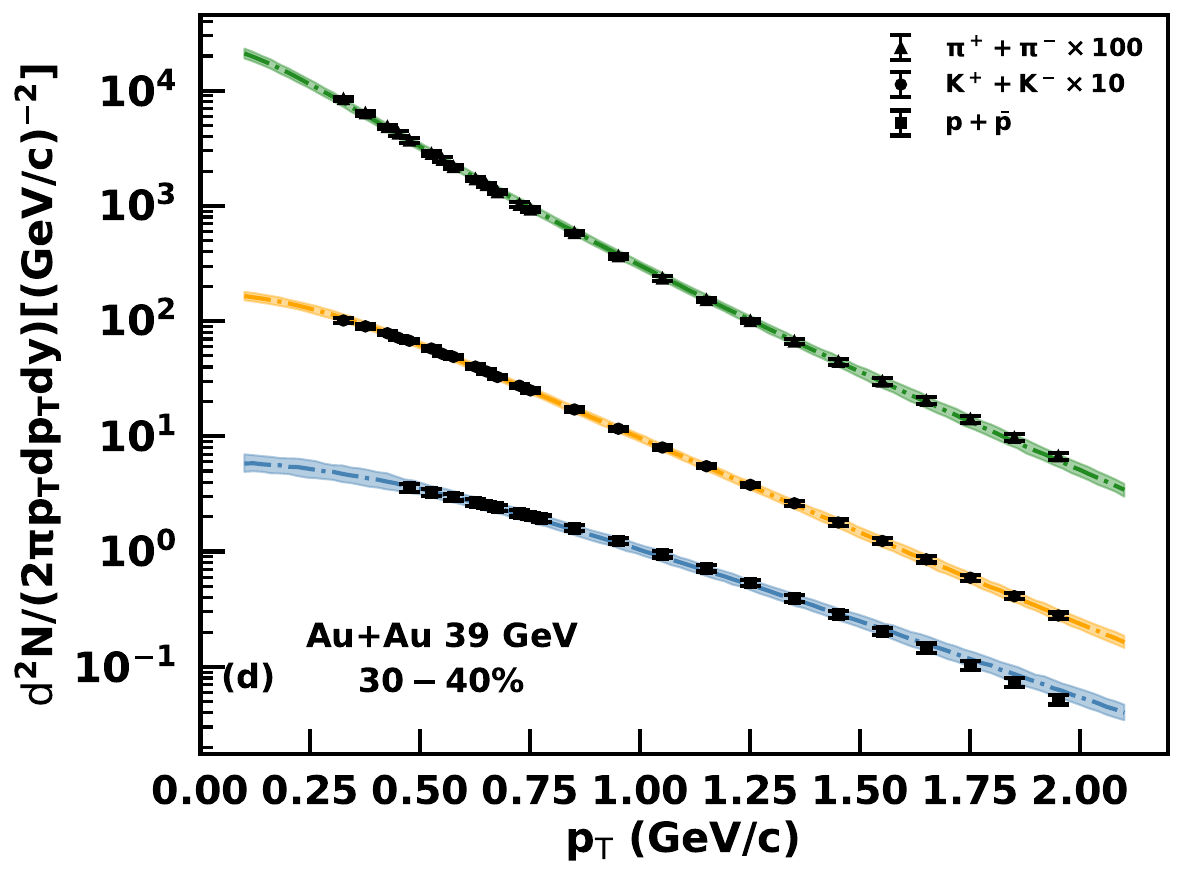}
    \end{overpic}
    }
    \hfill
   \caption{Transverse momentum spectra of $\pi^{+}+\pi^{-}$, $K^{+}+K^{-}$, and $p+\bar{p}$ in Au+Au collisions at $\sqrt{s_{\mathrm{NN}}} = 14.5$~GeV and $39$~GeV for two centrality classes (0–5$\%$ and 30–40$\%$) from the BGBW model compared with experimental data (markers) from STAR Collaboration \cite{STAR:2017sal,STAR:2019vcp}.}
  \label{fig:AuAu}
\end{figure*}
The particle $p_{\mathrm{T}}$ spectrum is written as follows:
\begin{eqnarray}\label{ptbgbw}
    \frac{d^2N}{2\pi p_Tdp_Tdy} &=& A\int_{0}^{R} m_{\mathrm{T}}I_0\left(\frac{p_{\mathrm{T}}\sinh\rho}{T_{kin}}\right)\nonumber\\
    &&\times K_1\left(\frac{m_{\mathrm{T}}\cosh\rho}{T_{kin}}\right)rdr,
\end{eqnarray}
where $A$ is a normalization factor for a given particle and is particle species-dependent. It is assumed that the particles are emitted from the same source, sharing the kinetic freeze-out temperature $T_{\text{kin}}$, the average transverse expansion velocity $\langle \beta_{\mathrm{T}} \rangle$, and the velocity profile exponent $n$. These six parameters, \(\{A_{\pi}, A_K, A_p, T_{\text{kin}}, \langle\beta_{\mathrm{T}}\rangle, n\}\), are extracted by performing simultaneous fits to the experimental data of $p_{\mathrm{T}}$ spectra of identified particles using Eq. (\ref{ptbgbw}).

\subsection{Bayesian inference approach}
To constrain the free parameters $\boldsymbol{\theta} = \{A_{\pi}, A_K, A_p, T_{\text{kin}}, \langle\beta_{\mathrm{T}}\rangle, n\}$ of the BGBW model, we employ a Bayesian inference approach by comparing the model predictions $y(\boldsymbol{\theta})$ with experimental $p_{\mathrm{T}}$ spectrum data $y_{\mathrm{exp}}$, following the methodology established in Ref.~\cite{Mantysaari:2025ltq}. According to Bayes' theorem, the posterior distribution of the parameters $P(\boldsymbol{\theta} | y_{\mathrm{exp}})$ is given by
\begin{equation}
    P(\boldsymbol{\theta} | y_{\mathrm{exp}}) = \frac{P(y_{\mathrm{exp}} | \boldsymbol{\theta}) \, P(\boldsymbol{\theta})}{P(y_{\mathrm{exp}})},
\end{equation}
where $P(y_{\mathrm{exp}} | \boldsymbol{\theta})$ is the likelihood, $P(\boldsymbol{\theta})$ is the prior, and $P(y_{\mathrm{exp}})$ is the evidence.

Following Ref.~\cite{Mantysaari:2025ltq}, Gaussian uncertainties are assumed for the likelihood function. Because the $p_{\mathrm{T}}$ spectra span over several orders of magnitude, the likelihood is constructed in logarithmic space, which weights the data points more evenly across the $p_{\mathrm{T}}$ range. It is given by
\begin{equation}
    P(y_{\mathrm{exp}} | \boldsymbol{\theta}) = \prod_{i=1}^{N_{\text{data}}} \frac{1}{\sqrt{2\pi}\,\varepsilon_i} \exp\left[ -\frac{1}{2} \left( \frac{\ln y_i(\boldsymbol{\theta}) - \ln y_{\mathrm{exp},i}}{\varepsilon_i} \right)^2 \right],
\end{equation}
where $y_i(\boldsymbol{\theta})$ is the model prediction for the $i$-th $p_{\mathrm{T}}$ bin from Eq.~(\ref{ptbgbw}), $y_{\mathrm{exp},i}$ is the corresponding experimental data point. Here $\varepsilon_i = \sigma_i / y_{\mathrm{exp},i}$ is the relative uncertainty of the $i$-th data point, with the total uncertainty defined as $
\sigma_i^2 = \sigma_{i,\mathrm{stat}}^2 + \sigma_{i,\mathrm{syst}}^2.
$ Since the full experimental covariance matrices are not available, statistical and
systematic uncertainties are added in quadrature and treated as uncorrelated, i.e., the relative uncertainties enter as a diagonal covariance in logarithmic space. This
approximation is regarded as a systematic limitation of the present analysis.

We use independent priors $P(\boldsymbol{\theta})$. Log-uniform priors are adopted for the normalization factors,
while uniform priors are used for  $\langle \beta_{\mathrm{T}} \rangle$, $T_{\text{kin}}$ and $n$.
\begin{itemize}
    \item $A$: the normalization factor for each particle species ($A_{\pi}$, $A_K$, $A_p$). A common log-uniform prior is adopted for all three species, with $\ln A$ sampled uniformly over the range $[1, 25]$ (equivalent to $p(A)\propto 1/A$). This range is wide enough to cover the absolute normalization of the $p_{\mathrm{T}}$ spectra for all collision systems, centralities, and energies considered, and does not favor any particular order of magnitude;
    \item Average transverse flow velocity $\langle \beta_{\mathrm{T}} \rangle$: $[0.10, 0.60]$ in Au+Au collisions, $[0.30, 0.70]$ in Pb+Pb collisions;
    \item $T_{\text{kin}} \in [0.02, 0.15]$~GeV;
    \item $n \in [0, 2.5]$. In addition, the physical constraint $\beta_s=\langle\beta_{\mathrm{T}}\rangle(n+2)/2<1$
is imposed during posterior sampling to ensure that the transverse flow velocity remains subluminal throughout the radial profile.
\end{itemize}
Posterior sampling is performed using the affine-invariant ensemble Markov chain Monte Carlo (MCMC) sampler implemented in the \texttt{emcee} package~\cite{Foreman-Mackey:2012any}. We use 100 walkers, discarding the first 500 steps as burn-in and retaining 1000 production steps. Convergence is assessed using the split-chain Gelman–Rubin $\hat R$ statistic, the integrated autocorrelation time $\tau$, the effective sample size (ESS), and the acceptance fraction: for all parameters $\hat R<1.01$, $\tau$ is much smaller than the chain length, the ESS exceeds $10^4$, and the mean acceptance fraction is around 0.5. The median values and 90$\%$ credible intervals of the posterior distributions are reported for each parameter.

\begin{figure*}[!ht]
     \centering
    \subfloat{
	\begin{overpic}[width=0.48\linewidth]{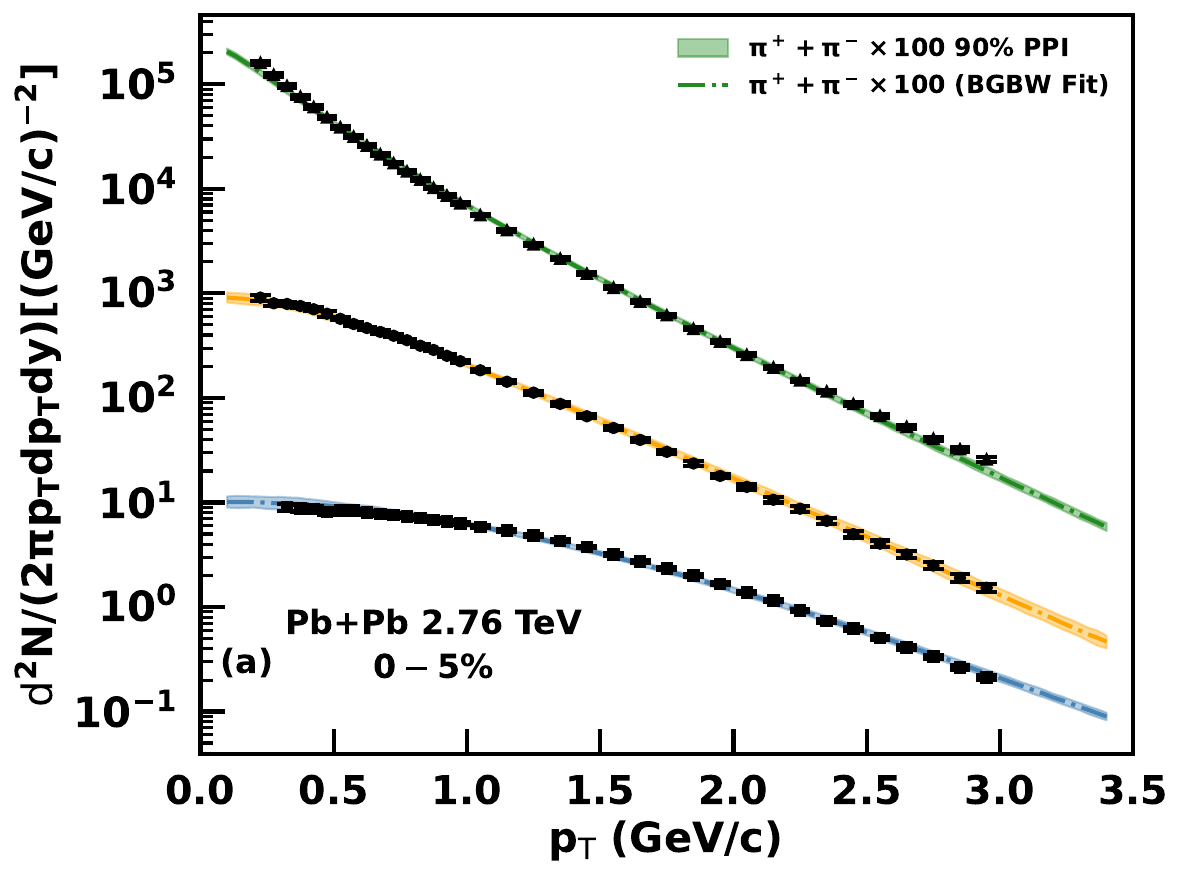}
    \end{overpic}
	}
    \hfill
     \subfloat{
	\begin{overpic}[width=0.48\linewidth]{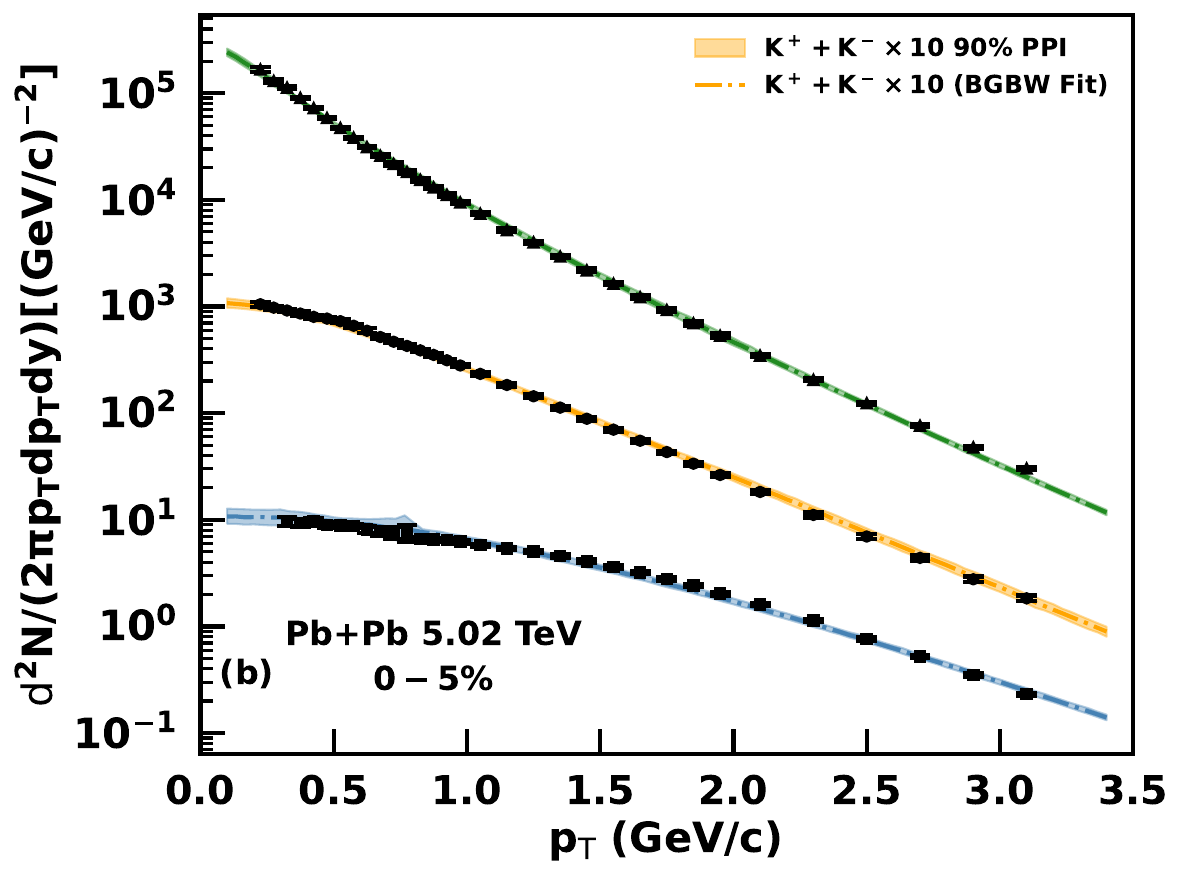}
    \end{overpic}
	}
    \hfill
    \subfloat{
	\begin{overpic}[width=0.48\linewidth]{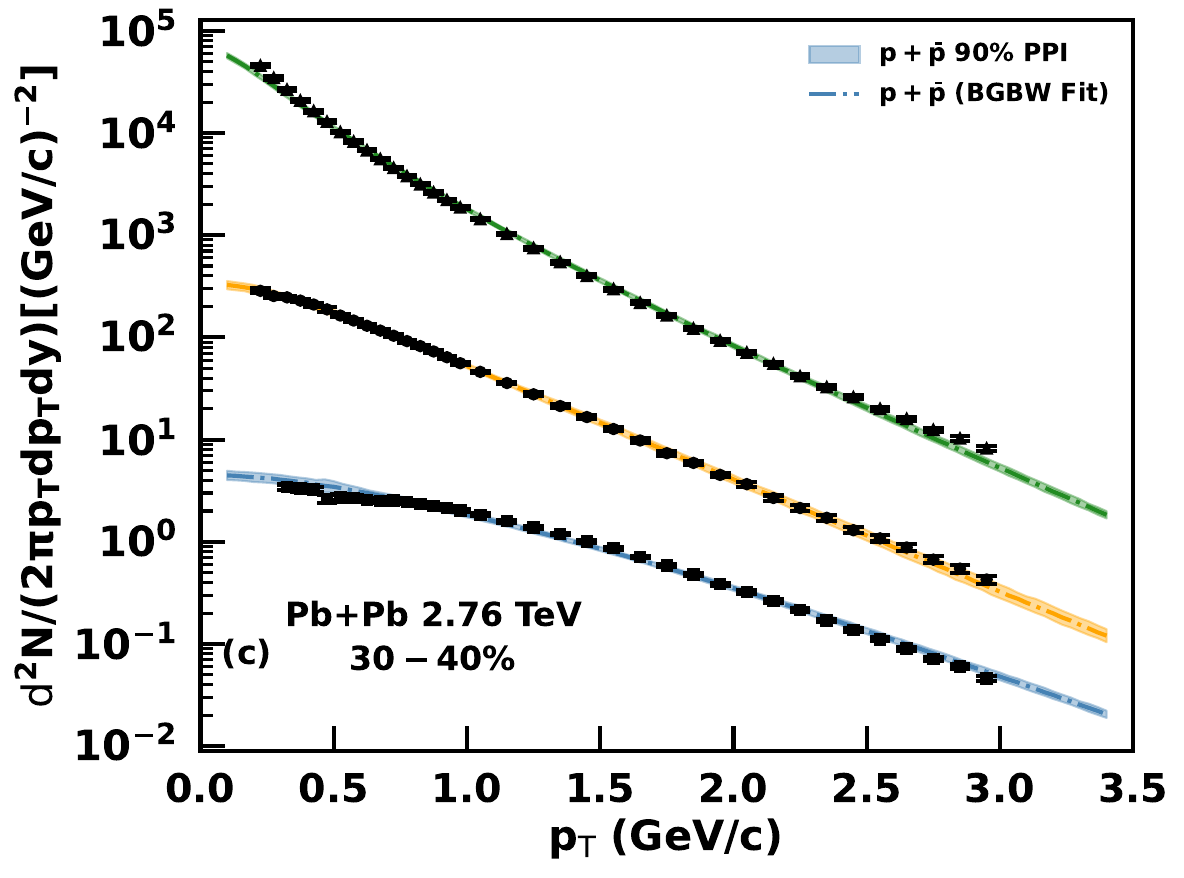}
    \end{overpic}
	}
    \hfill
    \subfloat{
	\begin{overpic}[width=0.48\linewidth]{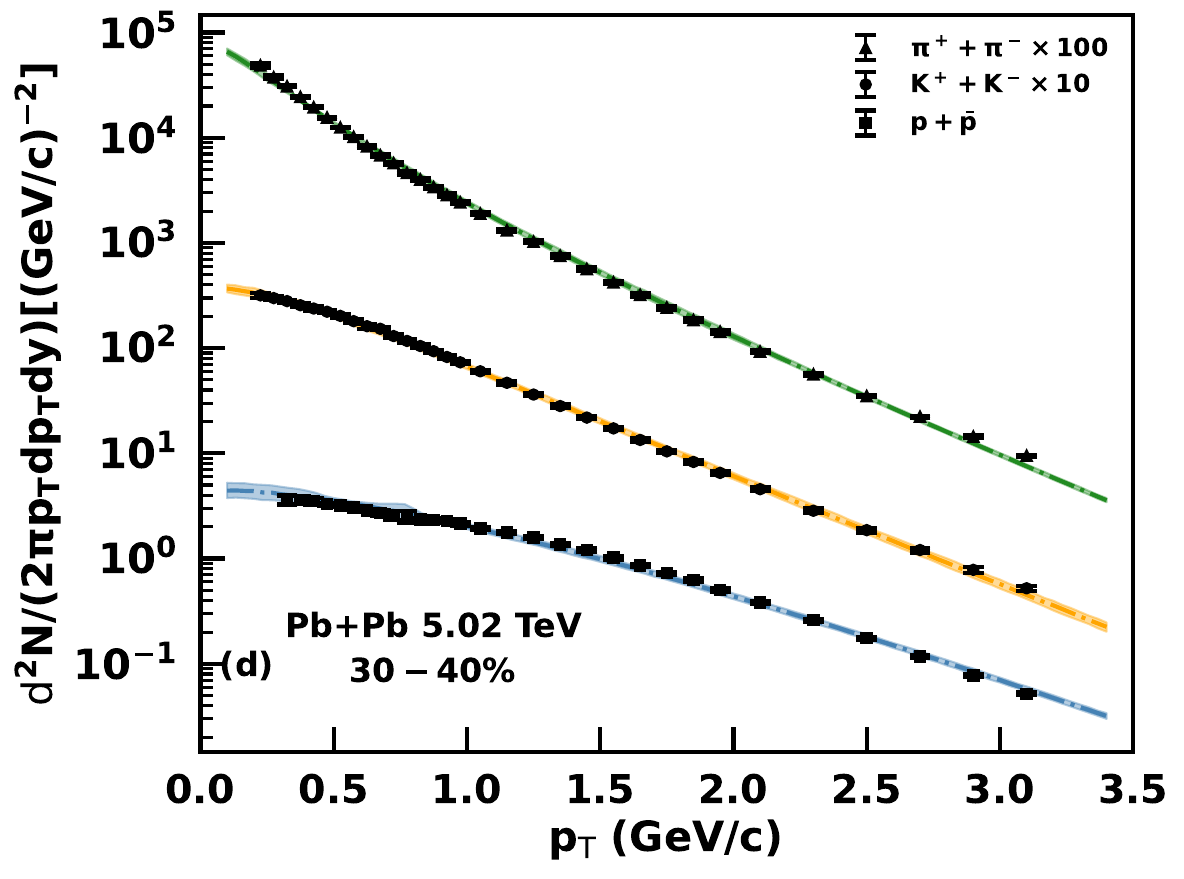}
    \end{overpic}
	}
    \hfill
    \subfloat{
	\begin{overpic}[width=0.48\linewidth]{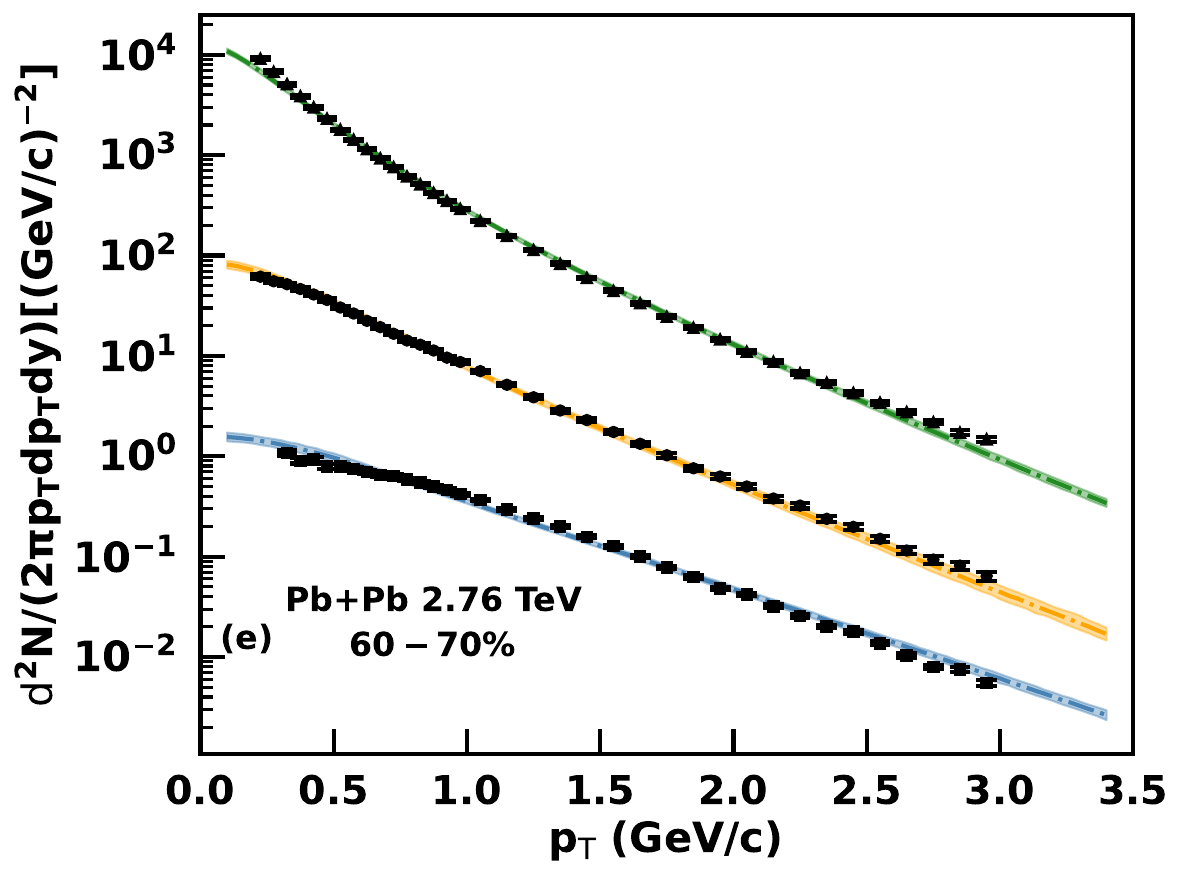}
    \end{overpic}
	}
    \hfill
    \subfloat{
	\begin{overpic}[width=0.48\linewidth]{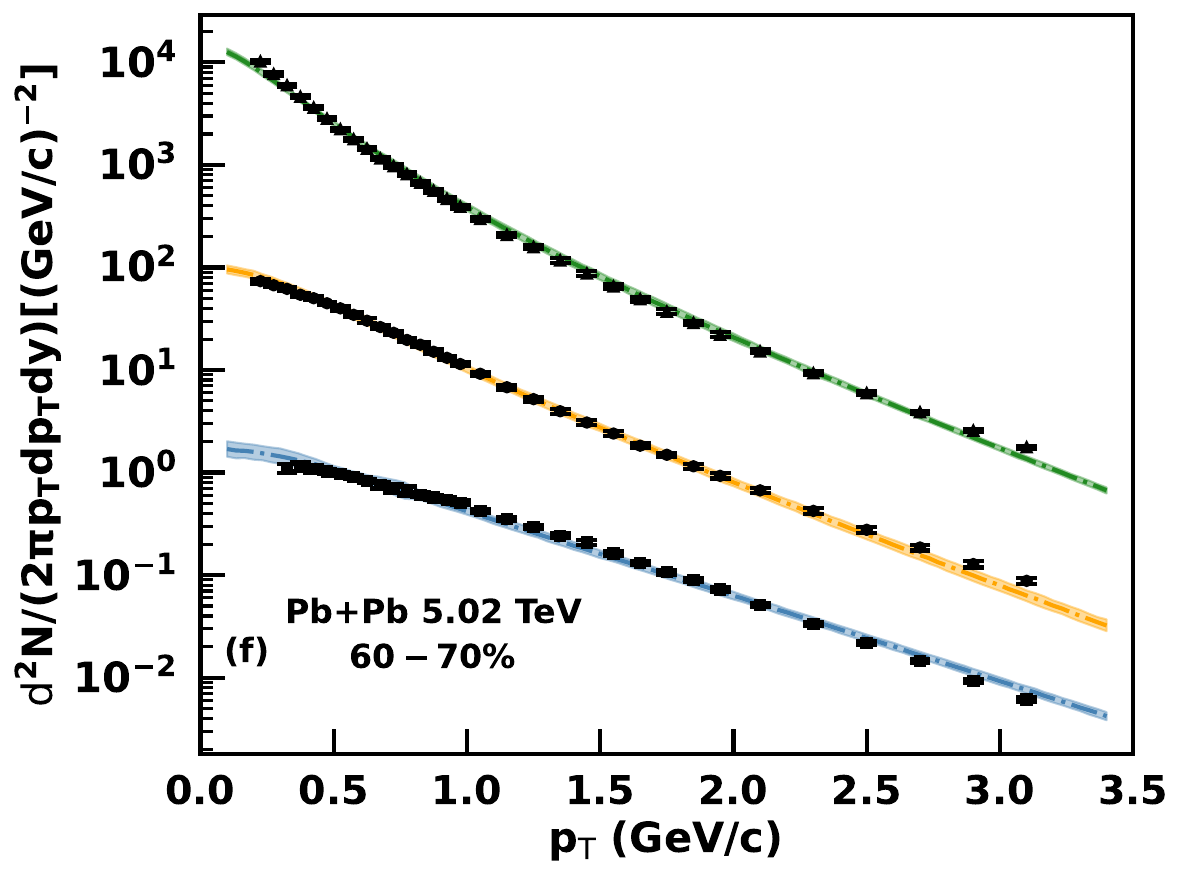}
    \end{overpic}
	}
    \hfill
   \caption{Transverse momentum spectra of $\pi^{+}+\pi^{-}$, $K^{+}+K^{-}$, and $p+\bar{p}$ in Pb+Pb collisions at $\sqrt{s_{\mathrm{NN}}} = 2.76$~TeV (left) and $5.02$~TeV (right) for three centrality classes (0–5$\%$, 30–40$\%$, 60–70$\%$) from the BGBW model compared with experimental data (markers) from ALICE Collaboration \cite{ALICE:2013mez,ALICE:2019hno}.}
  \label{fig:spectra}
\end{figure*}
\section{RESULTS AND DISCUSSION }\label{rad}

We revisit identified-particle $p_{\mathrm{T}}$ spectra using the BGBW model in a Bayesian inference framework. To demonstrate the advantages of Bayesian inference, we analyze the identified-particle $p_{\mathrm{T}}$ spectra data for a few collision centrality classes covering energies from the RHIC to LHC. For Au+Au collisions at $\sqrt{s_{\mathrm{NN}}}=7.7$–39 GeV, data from two centrality classes, 0–5$\%$ and 30–40$\%$, are analyzed. For Au+Au collisions at $\sqrt{s_{\mathrm{NN}}}=62.4$ and 200 GeV, data from the most central collisions are analyzed. At $\sqrt{s_{\mathrm{NN}}}=7.7$ and 11.5 GeV, only positively charged particles ($\pi^+$, $K^+$, $p$) are included, whereas for the other Au+Au collision energies, $\pi^{+}+\pi^{-}$, $K^{+}+K^{-}$, and $p+\bar{p}$ are analyzed. For Pb+Pb collisions at $\sqrt{s_{\mathrm{NN}}}=2.76$ and $5.02$ TeV, data from three centrality classes 0–5$\%$, 30–40$\%$ and 60–70$\%$ for $\pi^{+}+\pi^{-}$, $K^{+}+K^{-}$, and $p+\bar{p}$ are analyzed.

The $p_{\mathrm{T}}$ fit ranges for $p_{\mathrm{T}}$ spectra of identified particles in the Bayesian analysis are listed in Table \ref{tab:pT_ranges}. For comparison, the corresponding $p_{\mathrm{T}}$ fit ranges are also listed for the conventional BGBW model simultaneous fits using the minimum $\chi^2$/NDF. The lower bounds are determined by the available data. The upper bounds are also determined by the available data if the maximum $p_{\mathrm{T}}$ of the particle transverse momentum spectrum is less than 3 GeV/$c$. Otherwise the upper bounds are set to be 3 GeV/$c$. The extracted BGBW model parameters are listed in Table~\ref{tab1}.
\begin{table}[htbp]
  \centering
  \caption{Transverse momentum fit ranges for the BGBW model simultaneous fits to the $p_{\mathrm{T}}$ spectra of identified particles in Au+Au and Pb+Pb collisions.}
  \label{tab:pT_ranges}
  \begin{tabular}{l l c c c}
    \toprule
    System & Approach & $\pi$ (GeV/$c$) & $K$ (GeV/$c$) & $p$ (GeV/$c$) \\
    \midrule
    Au+Au & Conventional & 0.5–1.3   & 0.25–1.4 & 0.4–1.3   \\
          & This work     & 0.2–1.95  & 0.2–1.95 & 0.4–1.95  \\
    \cmidrule(lr){1-5}
    Pb+Pb & Conventional & 0.5–1.0   & 0.2–1.5  & 0.3–3.0   \\
          & This work      & 0.2–3.0   & 0.2–3.0  & 0.3–3.0 \\
    \bottomrule
  \end{tabular}
  \vspace{2pt}
\end{table}

\begin{table*}[ht]
	\centering
	\caption{The extracted BGBW model parameters using extended $p_{\mathrm{T}}$ fit ranges indicated in Table \ref{tab:pT_ranges} in Au+Au collisions at $\sqrt{s_{\mathrm{NN}}}= 7.7-200$ GeV from RHIC and Pb+Pb collisions at $\sqrt{s_{\mathrm{NN}}} = 2.76$~TeV and $5.02$~TeV from LHC.}
	\renewcommand{\arraystretch}{1.2} \tabcolsep 25pt% 
	\begin{tabular*}{1\textwidth}{ccccc}\toprule
    		$\sqrt{s_{\mathrm{NN}}}$ & Centrality& $\langle\beta_{\mathrm{T}}\rangle$ & $T_{\text{kin}}$ (GeV) & $n$ \\\hline
     7.7 GeV &$0-5\%$& $ 0.4605^{+0.0155}_{-0.0162}$& $0.1156^{+0.0042}_{-0.0038}$&$0.6347^{+0.1901}_{-0.2242}$ \\
	   &$30-40\%$ & $0.3669^{+0.0218}_{-0.0223}$& $0.1247^{+0.0048}_{-0.0045}$&$0.9166^{+0.3197}_{-0.3275}$ \\
\hline              
     11.5 GeV &$0-5\%$& $0.4418^{+0.0168}_{-0.0182}$& $0.1226^{+0.0048}_{-0.0043}$&$ 0.7023^{+0.2092}_{-0.2306}$ \\
	   &$30-40\%$ & $ 0.3586^{+0.0201}_{-0.0213}$& $0.1265^{+0.0048}_{-0.0045}$&$1.4043^{+0.2840}_{-0.2732}$ \\
\hline   
       14.5 GeV &$0-5\%$& $0.4166^{+0.0195}_{-0.0206}$& $ 0.1201^{+0.0043}_{-0.0040}$&$1.1683^{+0.2111}_{-0.2000}$ \\
	   &$30-40\%$ & $0.3387^{+0.0227}_{-0.0249}$& $0.1293^{+0.0045}_{-0.0044}$&$ 1.6772^{+0.3514}_{-0.3005}$ \\
\hline                 
      19.6 GeV &$0-5\%$& $ 0.4265^{+0.0136}_{-0.0136}$& $ 0.1180^{+0.0037}_{-0.0035}$&$ 1.2054^{+0.1395}_{-0.1451}$ \\
	   &$30-40\%$ & $0.3069^{+0.0189}_{-0.0216}$& $0.1349^{+0.0042}_{-0.0040}$&$2.0377^{+0.3566}_{-0.2987}$ \\
\hline      
      27 GeV &$0-5\%$& $0.4478^{+0.0132}_{-0.0140}$& $ 0.1189^{+0.0035}_{-0.0034}$&$1.0234^{+0.1324}_{-0.1364}$ \\
	   &$30-40\%$ & $  0.3446^{+0.0172}_{-0.0183}$& $0.1314^{+0.0041}_{-0.0040}$&$ 1.8476^{+0.2536}_{-0.2291}$ \\
\hline
     39 GeV &$0-5\%$& $0.4633^{+0.0119}_{-0.0122}$& $0.1192^{+0.0035}_{-0.0033}$&$1.0027^{+0.1230}_{-0.1191}$ \\
	   &$30-40\%$ & $0.3532^{+0.0165}_{-0.0167}$& $0.1330^{+0.0041}_{-0.0040}$&$1.8890^{+0.2291}_{-0.2158}$ \\
\hline
     62.4 GeV &$0-5\%$& $ 0.5470^{+0.0034}_{-0.0033}$& $ 0.1042^{+0.0018}_{-0.0018}$&$0.6403^{+0.0441}_{-0.0458}$ \\
\hline
     200 GeV &$0-12\%$& $0.4967^{+0.0206}_{-0.0207}$& $ 0.1156^{+0.0048}_{-0.0047}$&$ 1.1837^{+0.1637}_{-0.1520}$ \\     
\hline
       2.76 TeV &$0-5\%$& $0.6326^{+0.0038}_{-0.0038}$& $0.1036^{+0.0014}_{-0.0014}$&$0.7908^{+0.0216}_{-0.0218}$ \\
	   &$30-40\%$ & $0.5693^{+0.0045}_{-0.0045}$& $0.1164^{+0.0017}_{-0.0016}$&$1.0451^{+0.0288}_{-0.0281}$ \\
     &   $60-70\%$& $0.4312^{+0.0062}_{-0.0065} $& $0.1362^{+0.0021}_{-0.0021}$&$ 1.9151^{+0.0633}_{-0.0594}$\\
		\hline
       5.02 TeV  &$0-5\%$& $0.6423^{+0.0040}_{-0.0040}$& $0.1036^{+0.0015}_{-0.0014}$&$0.8029^{+0.0212}_{-0.0203}$\\
	  &$30-40\%$ & $0.5890^{+0.0041}_{-0.0042} $& $ 0.1184^{+0.0017}_{-0.0016}$&$ 0.9892^{+0.0255}_{-0.0254}$\\
   &   $60-70\%$& $0.4511^{+0.0075}_{-0.0075} $& $ 0.1402^{+0.0028}_{-0.0026}$&$ 1.7965^{+0.0678}_{-0.0670}$\\
   		\hline
	%\specialrule{0em}{2pt}{1pt}
	\bottomrule
	\end{tabular*}\label{tab1}
	\end{table*}

\begin{table*}[ht]
	\centering
	\caption{The extracted BGBW model parameters using different $p_{\mathrm{T}}$ fit ranges for the $0-5\%$ centrality class in Pb+Pb collisions at $\sqrt{s_{\mathrm{NN}}} = 5.02$~TeV.}
	\renewcommand{\arraystretch}{1.0} \tabcolsep 18pt% 
	\begin{tabular*}{0.9\textwidth}{ccccc}\toprule
    		 $p_{\mathrm{T}}$ range (GeV/$c$) &$\langle\beta_{\mathrm{T}}\rangle$ & $T_{\text{kin}}$ (GeV) & $n$ & Reference\\\hline
\makecell{$\pi$: 0.5--1 \\ $K$: 0.2--1.5 \\ $p$: 0.3--3}&
$0.663 \pm 0.003$ & $0.090 \pm 0.003$ & $0.735 \pm 0.013$ & \cite{ALICE:2019hno}\\
\hline
\makecell{$\pi$: 0.5--1 \\ $K$: 0.2--1.5 \\ $p$: 0.3--3} 
& $0.6617^{+0.0047}_{-0.0048}$ 
& $ 0.0921^{+0.0042}_{-0.0040}$ 
& $  0.7319^{+0.0207}_{-0.0202}$ & This work\\
\hline
\makecell{$\pi$: 0.2--1 \\ $K$: 0.2--1.5 \\ $p$: 0.3--3} 
& $0.6631^{+0.0044}_{-0.0043}$ 
& $ 0.0887^{+0.0026}_{-0.0025}$ 
& $ 0.7342^{+0.0198}_{-0.0197}$ & This work\\
\hline
\makecell{$\pi$: 0.5--3 \\ $K$: 0.2--3 \\ $p$: 0.3--3} 
& $ 0.6434^{+0.0041}_{-0.0041}$ 
& $0.1087^{+0.0017}_{-0.0017}$ 
& $ 0.7769^{+0.0219}_{-0.0217}$ & This work\\
\hline
\makecell{$\pi$: 0.2--3 \\ $K$: 0.2--3 \\ $p$: 0.3--3}
&$0.6423^{+0.0040}_{-0.0040}$& $0.1036^{+0.0015}_{-0.0014}$&$0.8029^{+0.0212}_{-0.0203}$ & This work\\     
	%\specialrule{0em}{2pt}{1pt}
	\bottomrule
	\end{tabular*}\label{tab2}
	\end{table*}

\begin{figure*}[ht]
   \centering
   \includegraphics[width=0.80\textwidth]{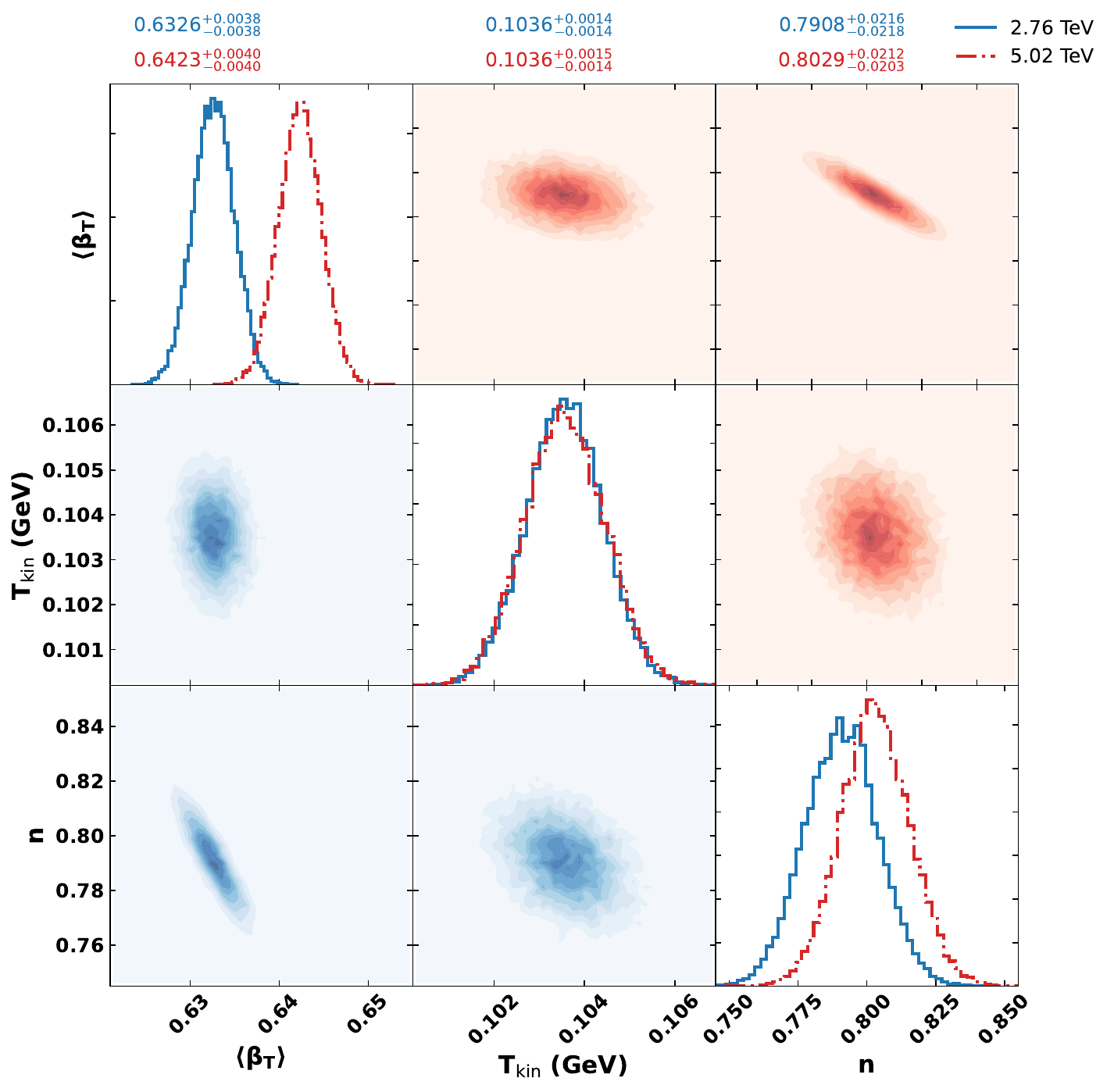}%
   \caption{Posterior distributions of the BGBW parameters $\langle\beta_{\mathrm{T}}\rangle$, $T_{\mathrm{kin}}$, and $n$ obtained from fitting $p_{\mathrm{T}}$ spectra of $\pi^{+}+\pi^{-}$, $K^{+}+K^{-}$, and $p+\bar{p}$ in Pb+Pb collisions at $\sqrt{s_{\mathrm{NN}}} = 2.76$~TeV (full, lower corner) and $5.02$~TeV (dash-dotted, upper corner) for the 0–5$\%$ centrality class. The numbers presented at the top of the figure are the median values along with their corresponding 90$\%$ credible intervals.}
  \label{fig:corner}
\end{figure*}

As representative examples, Fig.~\ref{fig:AuAu} presents the BGBW simultaneous posterior median predictions for the $p_{\mathrm{T}}$ spectra of identified particles $\pi^{+}+\pi^{-}$, $K^{+}+K^{-}$, and $p+\bar{p}$ using a Bayesian inference framework for 0-5$\%$ and 30-40$\%$ centrality classes in
Au+Au collisions at $\sqrt{s_\mathrm{NN}}$ = 14.5 and 39 GeV, compared with the experimental data.
Figure~\ref{fig:spectra} shows the same results for 0-5$\%$, 30-40$\%$, and 60-70$\%$ centrality classes in Pb+Pb collisions at $\sqrt{s_{\mathrm{NN}}} = 2.76$~TeV and $5.02$~TeV. In both figures, the experimental data are shown as solid markers with error bars: triangles for pions, circles for kaons, and squares for protons. The posterior median of the BGBW model predictions is shown as dash-dotted curves with colored bands indicating the $90\%$ posterior predictive intervals. For clarity, the pion and kaon spectra have been multiplied by factors of 100 and 10, respectively. The BGBW model provides a good description of the experimental data without enforcing the selected particle species-dependent $p_{\mathrm{T}}$ fit ranges, see Table \ref{tab:pT_ranges} for details. The simultaneous inference establishes a baseline for the thermal distribution under the BGBW model, thereby providing a foundation for subsequent studies on the low-$p_{\mathrm{T}}$ region.

Table~\ref{tab1} presents the BGBW model parameters extracted from the simultaneous fits to the $p_{\mathrm{T}}$ spectra of identified particles ($\pi$, $K$, and $p$) in Au+Au collisions at $\sqrt{s_{\mathrm{NN}}} = 7.7$–$200$ GeV from RHIC and Pb+Pb collisions at $\sqrt{s_{\mathrm{NN}}} = 2.76$ and $5.02$ TeV from LHC using Bayesian inference. The values are reported as medians with 90$\%$ credible intervals obtained from the posterior distributions. From Table~\ref{tab1}, clear centrality dependencies of the BGBW model parameters are observed for the collision energies and systems for which multiple centrality classes are analyzed. For both Au+Au and Pb+Pb collisions, the average radial flow velocity $\langle\beta_{\mathrm{T}}\rangle$ decreases from central to peripheral collisions. Correspondingly, the kinetic freeze-out temperature $T_{\text{kin}}$ increases from central to peripheral collisions, and the velocity profile exponent $n$ also increases significantly with increasing centrality. These trends are similar to the results obtained from the conventional BGBW model simultaneous fits to the data from RHIC and LHC in literature \cite{STAR:2017sal,STAR:2019vcp,ALICE:2013mez,ALICE:2019hno}. 

To examine how the selected particle species-dependent $p_{\mathrm{T}}$ fit ranges affect the values of extracted BGBW model parameters, we perform additional analyses by varying the analyzed $p_{\mathrm{T}}$ fit ranges of particle spectra, taking the most central Pb+Pb collisions at $\sqrt{s_{\mathrm{NN}}} = 5.02$~TeV as an example. The conventional BGBW model fitting ranges are $\pi$: 0.5--1 GeV/$c$, $K$: 0.2--1.5 GeV/$c$, and $p$: 0.3--3 GeV/$c$, same as in Table \ref{tab:pT_ranges}. We vary these fit ranges in two ways: one lowers the lower bound for pions from 0.5 to 0.2 GeV/$c$ which is the lowest $p_{\mathrm{T}}$ for the data; the other raises the upper bounds for pions and kaons to 3.0 GeV/$c$ which is the highest $p_{\mathrm{T}}$ analyzed in the conventional BGBW model studies. The corresponding extracted BGBW model parameters are summarized in Table~\ref{tab2}. Using the selected particle species-dependent $p_{\mathrm{T}}$ fit ranges, the extracted BGBW model parameters obtained in a Bayesian inference framework (the second row in Table~\ref{tab2}) are comparable to those from the conventional simultaneous fits~\cite{ALICE:2019hno}. Similar results are obtained for the case lowering the lower bound for pions (see the third row in Table~\ref{tab2}). When either raising the upper bounds for both the pions and kaons or both the lower and upper bounds are extended, $\langle\beta_{\mathrm{T}}\rangle$ decreases by about $\sim$3$\%$, $T_{\text{kin}}$ increases by about $\sim$15$\%$, and $n$ increases by about $\sim$9$\%$ compared with the conventional fit
in Ref. \cite{ALICE:2019hno}. Based on these results, it is plausible that we can simultaneously analyze the $p_{\mathrm{T}}$ spectra of identified particles ($\pi$, $K$, $p$) without enforcing the particle species-dependent $p_{\mathrm{T}}$ fit ranges using the BGBW model in a Bayesian inference framework. The extracted BGBW model parameters can also provide information about the collision system similar to those obtained from conventional analysis of the BGBW model.

To gain further insight into the extracted freeze-out parameters, we can examine their posterior distributions, which is one of the advantages of Bayesian analysis. Figure~\ref{fig:corner} shows corner plots for the average radial flow velocity $\langle\beta_{\mathrm{T}}\rangle$, the kinetic freeze-out temperature $T_{\text{kin}}$, and the velocity profile exponent $n$, which are obtained from Bayesian inference for Pb+Pb collisions at $\sqrt{s_{\mathrm{NN}}} = 2.76$~TeV and $5.02$~TeV, respectively, as examples. The results for the 0–5$\%$ centrality class are presented. The posterior distributions of the three parameters are unimodal and well constrained, indicating that the data provide sufficient information to determine the freeze-out parameters. The relations between the BGBW model parameters are similar for Pb+Pb collisions at $\sqrt{s_{\mathrm{NN}}} = 2.76$~TeV and $5.02$~TeV. This information is presented more transparently through the posterior correlations. As shown in Fig.~\ref{fig:corner}, no clear correlation between $T_{\text{kin}}$ and $\langle\beta_{\mathrm{T}}\rangle$ is observed. A clear anticorrelation is observed between the velocity profile exponent $n$ and $\langle\beta_{\mathrm{T}}\rangle$. This anticorrelation can be attributed to the relation between the surface flow velocity $\beta_s = \langle\beta_{\mathrm{T}}\rangle(n+2)/2$ and the high-$p_{\mathrm{T}}$ particles in the transverse momentum spectrum. 
A weak anticorrelation between $n$ and $T_{\text{kin}}$ is observed. 

\section{CONCLUSIONS}\label{con}
In this work, we have revisited the transverse momentum ($p_{\mathrm{T}}$) spectra of identified particles, i.e., pions, kaons, and protons, at midrapidity in Au+Au collisions at $\sqrt{s_{\mathrm{NN}}} = 7.7$–$200$ GeV and in Pb+Pb collisions at $\sqrt{s_{\mathrm{NN}}} = 2.76$ and $5.02$ TeV using the Boltzmann–Gibbs blast-wave model in a Bayesian inference framework. We demonstrate that the BGBW model can simultaneously describe the $p_{\mathrm{T}}$ spectra of identified particles well without enforcing the particle species-dependent $p_{\mathrm{T}}$ fit ranges -- a practice that was followed in conventional BGBW model analyses of the $p_{\mathrm{T}}$ spectra of identified particles to achieve reasonably good simultaneous fits. It is shown that the extracted BGBW model parameters remain broadly consistent with those from conventional simultaneous fits, while the extension of the fit range leads to moderate changes in some parameters. With our systematic analyses, the behaviors of the parameters as functions of collision energy and/or centrality, as well as their correlations, are similar to those from the conventional approach. Bayesian analysis also provides well-constrained posterior distributions for the extracted BGBW model parameters and quantifies the correlations among the model parameters. Therefore, the BGBW model in a Bayesian inference framework is suggested for use in future data analyses to simultaneously describe the $p_{\mathrm{T}}$ spectra of identified particles and extract the relevant information about the collision system.

\section*{Acknowledgement}
The authors thank Prof. Wenbin Zhao for helpful discussions. 

\bibliography{reference.bib}

\end{document}